\documentclass[prd,a4paper,showpacs,preprint,byrevtex]{revtex4}
\usepackage{graphicx}
\usepackage{dcolumn}
\usepackage{amsmath}
\usepackage{array}
\usepackage{bm}
\usepackage{textcomp}
\begin{document}

\title{Hybrid mesons with auxiliary fields}

\author{Fabien \surname{Buisseret}}
\thanks{FNRS Research Fellow}
\email[E-mail: ]{fabien.buisseret@umh.ac.be}
\author{Vincent \surname{Mathieu}}
\thanks{IISN Scientific Research Worker}
\email[E-mail: ]{vincent.mathieu@umh.ac.be}
\affiliation{Groupe de Physique Nucl\'{e}aire Th\'{e}orique,
Universit\'{e} de Mons-Hainaut,
Acad\'{e}mie universitaire Wallonie-Bruxelles,
Place du Parc 20, BE-7000 Mons, Belgium}

\date{\today}

\begin{abstract}
Hybrid mesons are exotic mesons in which the color field is not in the ground state. Their understanding deserves interest from a theoretical point of view, because it is intimately related to nonperturbative aspects of QCD. Moreover, it seems that some recently detected particles, such as the $\pi_1(1600)$ and the $Y(4260)$, are serious hybrid candidates. In this work, we investigate the description of such exotic hadrons by applying the auxiliary fields technique (also known as the einbein field method) to the widely used spinless Salpeter Hamiltonian with appropriate linear confinement. Instead of the usual numerical resolution, this technique allows to find simplified analytical mass spectra and wave functions of the Hamiltonian, which still lead to reliable qualitative predictions. We analyse and compare two different descriptions of hybrid mesons, namely a two-body $q\bar q$ system with an excited flux tube, or a three-body $q\bar q g$ system. We also compute the masses of the $1^{-+}$ hybrids. Our results are shown to be in satisfactory agreement with lattice QCD and other effective models.
\end{abstract}


\maketitle

\section{Introduction}
The study of hybrid mesons is an active domain in theoretical as well as in experimental particle physics. From a theoretical point of view, these particles are interpreted as mesons in which the color field is in an excited state. Clearly, this problem is related to fundamental aspects of QCD, such as its nonperturbative nature. Numerous lattice QCD calculations have been devoted to the study of hybrid mesons, in particular to the properties of the $1^{-+}$ state, which is the lightest hybrid with exotic quantum numbers (see Ref.~\cite{Nel02} for a review, and Ref.~\cite{Liu06} for more recent references). On the experimental side, we can mention the recently observed $\pi_1(1600)$ \cite{Ada98} and $Y(4260)$ \cite{aub05}, which could be interpreted as a $1^{-+}$ $n\bar n$ hybrid and a $1^{--}$ $c\bar c$ hybrid respectively \cite{Klem04}.  
\par Apart from lattice QCD, hybrid mesons have been studied with effective models for a long time. For example, we can quote the flux tube model \cite{flux}, models with constituent gluons \cite{constg}, or the MIT bag model \cite{bag}. In potential approaches, to which our paper is devoted, there are two main models. In the first one, the quark and the antiquark are linked by a string, or flux tube, which simulates the exchange of gluons responsible for the confinement. If the string is in the ground state, it reduces to the usual linear confinement potential for heavy quarks, and to a more general flux tube model for light quarks, where the dynamics of the string cannot be neglected \cite{laco89,Ols94}. In this stringy picture, it is possible for the flux tube to fluctuate, and thus to be in an excited state. These string excitations are analog to the gluon field excitations in full QCD. They have been studied for example in Refs.~\cite{Allen:1998wp,luscher}. The second approach is to suppose that the hybrid meson is a three-body system, formed of a quark, an antiquark, and a constituent gluon, which represents the gluonic excitation. Two fundamental strings then link the gluon to the quark and to the antiquark. This picture has been studied in Ref.~\cite{constg}, but also in more recent works \cite{Szczepaniak:2005xi,Szczepaniak:2006nx,Mathieu:2005wc,abre05,Kalashnikova:2002tg}. 

Nowadays, the spinless Salpeter Hamiltonian (SSH) with a linear confinement is a widely used and successful framework to compute hadron spectra in potential models (see previous references). Since its kinetic operator is semi-relativistic, most of the calculations have to be numerical. However, the auxiliary field (AF) technique allows to greatly simplify the calculations \cite{Sem03} and, as we will see, to find analytical solutions to this problem. Even if they are approximated, they lead to conclusions which are qualitatively in agreement with well-known experimental facts. In particular, Regge trajectories are easily obtained for light mesons by using AF \cite{qse}. Our purpose is to apply here this formalism in order to get informations about hybrid mesons. This formalism, also known as the einbein field method, has been applied to quark-antiquark two-body systems in Ref. \cite{Kalashnikova:1996pu,Dirac}. This method can be generalized for the case of spinning particles \cite{Brink:1976uf}.

\par Our paper is organized as follows. In Sec.~\ref{mesonic}, we solve the SSH in the case of a two-body problem. Although this simple case is relatively well-known, it will allow us to introduce the AF formalism, and to observe that it leads to correct predictions. Then we present in Sec.~\ref{hyb_ex} the description of a hybrid meson in terms of a $q\bar q$ system in which the flux tube is excited. An other possible approach is to see the hybrid mesons as a three-body system made of a quark-antiquark pair and a constituent gluon. This case is studied in Sec.~\ref{hyb_g}. As a result, we are able to compute the effective $q\bar q$ potential in both approaches. These potentials can be compared to the predictions of other effective models and of lattice QCD calculations. It is done in Sec.~\ref{eff_pot}. Finally, we compute the spectrum of the $1^{-+}$ hybrids in Sec.~\ref{spectrum}, and we sum up our results in Sec.~\ref{conclu}. 

\section{The two-body problem}\label{mesonic}

\subsection{Mass formula and wave function}
The SSH for a system made of two hadrons interacting through a linear confinement is given by
\begin{equation}\label{ham1}
H=\sqrt{\bm{p}^{\, 2}_1+m^2_1}+\sqrt{\bm{p}^{\, 2}_2+m^2_2}+ar.	
\end{equation}
Let us now introduce three AF (or einbein fields): Two for the quarks, denoted $\mu_i$, and one for the potential, $\nu$. Hamiltonian (\ref{ham1}) then becomes
\begin{equation}\label{ham2}
H(\mu_i,\nu)=\frac{\bm{p}^{\, 2}_1+m^2_1}{2\mu_1}+\frac{\mu_1}{2}+	\frac{\bm{p}^{\, 2}_2+m^2_2}{2\mu_2}+\frac{\mu_2}{2}+\frac{a^2 r^2}{2\nu}+\frac{\nu}{2}.
\end{equation}
The AF were introduced to get rid of the square roots in H. Although being formally simpler, $H(\mu_i,\nu)$ is equivalent to $H$ up to the elimination of the auxiliary fields thanks to the constraints
\begin{subequations}\label{elim}
\begin{eqnarray}
	\delta_{\mu_i}H(\mu_i,\nu)&=&0\ \Rightarrow\ \mu_{i0}=\sqrt{\bm{p}^{\, 2}_i+m^2_i},\\
	\delta_{\nu}H(\mu_i,\nu)&= &0\ \Rightarrow\ \nu_{0}=ar.
\end{eqnarray}
\end{subequations}
It is worth mentioning that $\left\langle \mu_{i0}\right\rangle$ can be seen as a dynamical mass of the quark whose current mass is $m_i$ \cite{Fab1}. Moreover, the Hamiltonian (\ref{ham2}) can be compared to the one of the rotating string model (RSM)  \cite{rsm}. This is an effective meson model derived from the QCD Lagrangian, in which the quark and the antiquark are linked by a Nambu-Goto string simulating the confining interaction. The RSM Hamiltonian reads, in the center of mass frame,
\begin{eqnarray}
\label{QCD_eq1}
H^{RSM}(\mu_i,\nu)=\frac{1}{2}\left\{\frac{p^{2}_{r}+m^{2}_{1}}{\mu_{1}}+\frac{p^{2}_{r}
+m^{2}_{2}}{\mu_{2}}+\mu_{1}+\mu_{2}+ a^{2}r^{2}\int^{1}_{0}\frac{d\beta}{
\nu}+ \int^{1}_{0}d\beta\nu+\frac{\vec{L}^{\, 2}}{a_3 r^{2}}\right\},
\end{eqnarray} 
with
\begin{equation}
	a_3=\mu_{1}(1-\zeta)^{2}+\mu_{2}\zeta^{2}+\int^{1}_{0} d\beta\,
(\beta-\zeta)^{2}\, \nu.
\end{equation}

The parameter $\beta$ labels the points of the string. $\zeta$ defines the position of the center of mass on the string, and $\mu_i$, $\nu$ are the AF. In this framework, $\nu$ is seen as the dynamical energy of the string whose ``static" energy is $ar$. The complexity of (\ref{QCD_eq1}) is due to the fact that the string contributes to the total angular momentum $\vec{L}$. If we neglect the dynamical effects of the string, which are in fact sufficiently small to be be treated in perturbation (see Ref~\cite{qse}), Hamiltonian (\ref{QCD_eq1}) becomes
\begin{equation}\label{ham4}
	 H(\mu_i,\nu)=\frac{\bm{p}^{\, 2}_1+m^2_1}{2\mu_1}+\frac{\mu_1}{2}+	\frac{\bm{p}^{\, 2}_2+m^2_2}{2\mu_2}+\frac{\mu_2}{2}+\frac{a^2 r^2}{2\nu}+\frac{\nu}{2},
\end{equation}
 which is precisely our SSH (\ref{ham2}). 
 \par We can observe from the relations (\ref{elim}) that the AF are, strictly speaking, operators. However, the calculations are considerably simplified if one considers them as real numbers. The elimination of the AF is then finally achieved by minimising the masses with respect to them \cite{Sem03}. This procedure leads to a spectrum which is an upper bound of the ``true spectrum" (computed without AF), the differences being about $10\%$ \cite{Fab3}. In our case, the SSH turns out to be a simple nonrelativistic harmonic oscillator (\ref{ham4}). Its mass spectrum and wave functions are thus readily computed. They read 
\begin{equation}
M(\mu_i,\nu)=\omega(2n+\ell+3/2)+\frac{m^2_1}{2\mu_1}+\frac{m^2_2}{2\mu_2}+\frac{\mu_1+\mu_2+\nu}{2},
\end{equation}  
\begin{equation}\label{phico}
\psi=\phi_{n,\ell}(r)Y^{m}_{\ell}(
\theta,\varphi),
\end{equation}
with
\begin{equation}
\omega=	\sqrt{a^2/\tilde{\mu}\nu},\quad \beta=\sqrt{\tilde{\mu} a^2/\nu},\quad \tilde\mu=\frac{\mu_1\mu_2}{\mu_1+\mu_2}.
\end{equation}
$Y^{m}_{\ell}$ are the spherical harmonics, and
\begin{eqnarray}\label{osc3d}
\phi_{n,\ell}=\beta^{\frac{1}{2}(\ell+3/2)}\sqrt{2n!/
\Gamma(n+\ell+3/2)}\ r^{\ell}\, e^{-\beta r^{2}/2}\,  L^{\ell+
\frac{1}{2}}_{n}(\beta r^{2})
\end{eqnarray}
is a properly normalised radial eigenfunction of the three dimensional
harmonic oscillator~\cite{Brau}.
$L^{\alpha}_{n}$ are the Laguerre polynomials.
\par $\nu$ is eliminated by demanding $\delta_\nu M(\mu_i,\nu)=0$, which leads to 
\begin{equation}
	\nu_0=(a^2/\tilde{\mu})^{1/3}(2n+\ell+3/2)^{2/3},
\end{equation}
\begin{eqnarray}\label{mass2_1}	M(\mu_i,\nu_0)&=&\frac{3}{2}\left(\frac{a^2}{\tilde{\mu}}\right)^{1/3}(2n+\ell+3/2)^{2/3}+\frac{m^2_1}{2\mu_1}+\frac{m^2_2}{2\mu_2}+\frac{\mu_1+\mu_2}{2}.
\end{eqnarray}
\par The remaining AF, $\mu_i$, cannot be analytically eliminated in general from the condition $\delta_{\mu_i}M(\mu_i,\nu_0)=0$. We will only consider three relevant special cases. Firstly, if the two bodies have a large mass, we can set $\mu_i=m_i$ because the dynamical effects will be very small, and we obtain 
\begin{equation}\label{mass2_hh}
	M_{hh}=\frac{3}{2}\left(\frac{2a^2}{\tilde{m}}\right)^{1/3}(2n+\ell+3/2)^{2/3}+m_1+m_2,
\end{equation}
with $\tilde m$ the reduced mass. The mass formula (\ref{mass2_hh}) is valid for example for a meson formed of two heavy quarks. It is equal at large $\ell$ to the corresponding classical solution of the relativistic flux tube model \cite{Ols94}. This phenomenological model is in fact classically equivalent to the RSM if the auxiliary fields are properly eliminated \cite{Fab1}. Moreover, at large angular momentum, $M_{hh}\propto\ell^{2/3}$ in qualitative agreement with the experimental data \cite{Ols94}. Secondly, if $m_i=0$, as it is the case for light mesons and glueballs formed of two gluons, we can compute that 
\begin{equation}
	\mu_{1,0}=\mu_{2,0}=\mu_0=\left(\frac{a}{2}\right)^{1/2}(2n+\ell+3/2)^{1/2},
\end{equation}
\begin{equation}\label{mass2_3}
	M_{ll}=4 \left(\frac{a}{2}\right)^{1/2}(2n+\ell+3/2)^{1/2}=4\mu_0,
\end{equation}
as it is expected from the relativistic virial theorem \cite{RVT}. Squaring (\ref{mass2_3}), we get
\begin{equation}\label{mass2_4}
	M^2_{ll}=8a(2n+\ell+3/2).
\end{equation}
 When $\ell$ is large, it appears that the square mass increases linearly with $\ell$. Thus, our solution reproduces the Regge trajectories, which are the best known experimental fact concerning the light meson spectroscopy. The Regge slope is given by $8a$, which is in agreement with a recent calculation of the glueball spectrum with the RSM \cite{Simo1}. However, it is larger than the prediction of the relativistic flux tube, that is $2\pi a$ \cite{laco89}. This is in fact related to the AF technique itself, and more precisely to the number of AF which have to be introduced, as explained in Appendix \ref{AFM}. With $a\approx 0.2$\, GeV$^2$, a mass formula such as (\ref{mass2_4}) is able to correctly reproduce the experimental Regge slope of the mesons \cite{laco89}.
 Finally if, say, $m_1=0$ and $m_2$ is large, we can find 
\begin{equation}\label{masshl}
	(M_{hl}-m_2)^2=4a(2n+\ell+3/2).
\end{equation}
 The Regge slope for a meson formed of a light and a heavy quark is thus the half of the one for two light quarks, as it was shown is Ref.~\cite{Ols94}, in agreement with experimental observations.

\section{Hybrid meson and the excited flux tube}\label{hyb_ex}

If the color field is in the ground state, it is generally accepted that the potential between the quark and the antiquark in a meson is mainly compatible with a funnel potential,
\begin{equation}\label{pot_qq}
	V_{q\bar q}(r) = ar - \frac{4\alpha_S}{3r}, 
\end{equation}
where $\alpha_S$ is the strong coupling constant. The $ar$ part is pure flux tube, thus pure confinement, while the Coulomb term comes from the one gluon exchange process (OGE). The spectrum obtained with (\ref{pot_qq}) is in good agreement with experimental data for the light and heavy mesons \cite{truc}, but also with lattice QCD calculations \cite{Koma}. Typical values for the parameters fitting these lattice QCD data are $a = 0.2$ GeV$^2$, and $\alpha_S =0.2- 0.3$. 
\par In a hybrid meson, we have to wonder about how this potential will be modified. A well-known approach, based on the computation of the flux tube fluctuations at the quantum level, leads to the so-called L\"uscher term. In this approach, the potential between two fixed quarks is given by \cite{luscher}
\begin{equation}\label{pot_lu}
	V_{q\bar q}(r)=ar+\frac{\pi}{r}\left(N-\frac{1}{12}\right),
\end{equation}
where $N$ is the excitation number of the string. For $N=0$, we recover the Funnel potential(\ref{pot_qq}) with formally $\alpha_S=\pi/16\approx0.2$. This corresponds more or less to the usual value. For $N>0$, the short-range term becomes repulsive, and this potential should become applicable to heavy hybrid mesons. 
\par Another formalism has also been developed in Ref.~\cite{Allen:1998wp} to treat the excitations of the flux tube. It is based on less conventional approaches to string theory, and leads to 
\begin{equation}\label{pot_ols}
	V_{q\bar q}(r)=\sqrt{a^2r^2+2\pi a N}+\frac{\alpha_S}{6r},
\end{equation}
where the short-range term is not due to the string, but again to OGE, with the $q\bar q$ pair in an octet. Let us note that for large $r$, $\sqrt{a^2r^2+2\pi a N}\approx ar+\pi N/r$, and we approximately recover the L\"uscher term. As we can consider that the short-range term can be added in perturbation in a first approach, the unperturbed spectrum of the potential (\ref{pot_lu}) will be defined by the mass formula we derived in Sec. \ref{mesonic} from Eq.~(\ref{mass2_1}). However, the confining part of potential (\ref{pot_ols}) being different, it will affect the mass spectrum, even at the unperturbed level. If $N\neq 0$, the calculations cannot be analytically performed, except for two heavy quarks. In this case, one can readily obtain a sort of Regge trajectory with respect to $N$,
\begin{equation}\label{regge_hyb}
	(M_{hh}-2m)^2\approx 2\pi a N.
\end{equation}

\section{Hybrid mesons with constituent gluons}\label{hyb_g}

In this picture, it is assumed that the excitations of the gluon field can be described by the potential created by a constituent gluon. The quark-antiquark pair is thus in a color octet in order for the hybrid to be a colorless object. Assuming the Casimir scaling hypothesis, which seems to be confirmed by several models \cite{scaling}, it can be shown that the confinement is no more a Y-junction like in a baryon but two fundamental strings linking each quarks to the gluon \cite{Mathieu:2005wc}. Neglecting all the short-range interactions, the three-body SSH is thus 
\begin{equation}\label{mainH}
	H_0=\sum_{i=q,\bar q,g}\sqrt{\bm p^2_i+m^2_i}+a|\bm x_g-\bm x_q|+a|\bm x_g-\bm x_{\bar q} |.
\end{equation}

Let us now consider a hybrid meson in which the quark and the antiquark have the same mass, and are assumed to be static like in lattice QCD. Then, all the properties of the hybrid meson should depend only on the quark-antiquark separation and on the quantum numbers of the gluon. If we define $\bm x_q-\bm x_{\bar q}=\bm R$, and $\bm r=-\bm R/2+\bm x_g-\bm x_{\bar q}=\bm R/2+\bm x_g-\bm x_{\bar q}$, the potential of the strings is proportional to $|\bm r+\bm R/2| + |\bm r-\bm R/2|$.
As this expression is not useful in a practical computation, we will use the Born-Oppenheimer approximation
\begin{equation}\label{approx}
	|\bm r+\bm R/2| + |\bm r-\bm R/2|\approx 2\sqrt{r^2+R^2/4},
\end{equation}
valid for $\cos (\bm r,\bm R)\ll r/R + R/4r$. This upper bound is always greater than $1$ (the mimimum being reached for $2r=R$). Actually, our assumption says that the gluon lies on the symmetry plane of the $q\bar q$ pair. The Hamiltonian of the system for static quarks ($\bm p^2_{q, \bar q}=0$) reads then
\begin{equation}\label{Hqqg1}
	H_0 = \sqrt{\bm p^2} + 2a\sqrt{\bm r^2+R^2/4}+2m.
\end{equation}
The string tension $a$ is the same than in the meson case. We assumed an equal mass $m$ for the quark and the antiquark and a vanishing current mass for the gluon.  

The eigenvalues of the Hamiltonian \eqref{Hqqg1} can be found for any value of the quark-antiquark separation by introducing two AF, $\mu$ being again the constituent mass of the gluon and $\nu$ the energy of the strings. With these two fields, Eq.~\eqref{Hqqg1} becomes a harmonic oscillator Hamiltonian,
\begin{equation}\label{Hqqg2}
	H_0(\mu,\nu) = \frac{\bm p^2}{2\mu}+\frac{\mu}{2} + \frac{2a^2 r^2}{\nu}+\frac{a^2R^2}{2\nu}+\frac{\nu}{2}+2m.
\end{equation}
Its eigenvalues
\begin{equation}\label{energ_h1}
	E_0(\mu,\nu)-2m = \frac{2a(2n+\ell+3/2)}{\sqrt{\mu\nu}}+\frac{\mu+\nu}{2}+\frac{a^2R^2}{2\nu},
\end{equation}
depend on $R$, and on the quantum numbers $n$ and $\ell$ of the gluon. Their eigenfunctions are given by (\ref{phico}), with 
\begin{equation}\label{betadef3}
\beta=2a\sqrt{\mu/\nu}.
\end{equation}
The constraints $\delta_{\mu,\nu}E_0(\mu,\nu)=0$ lead to 
\begin{equation}\label{mudef}
	\nu_0=k^2\mu_0^{-3},
\end{equation}
\begin{equation}\label{condi}
	a^2R^2 \mu^6_0 + k^2\mu^4_0- k^4=0,
\end{equation}
with 
\begin{equation}
k=2a(2n+\ell+3/2).	
\end{equation}
Equation (\ref{mudef}) defines $\nu_0$ in terms of $\mu_0$, and this last AF is found to be a solution of Eq.~(\ref{condi}). This equation can be solved for any value of $R$ thanks to the Cardan method (see Appendix B).

\par Two limit cases are interesting. The first one is the limit of large $R$. Then, $\mu^3\approx k^{2}/aR$, $\nu\approx aR$, and 
\begin{equation}
	E_0-2m\approx aR.
\end{equation}
The effective current mass for the gluon decreases with the interquark distance. For large quark separation, the energy is only given by the flux tube and the potential energy is the expected linear confinement. Secondly, if $R=0$, we have $\mu=\nu=\sqrt{k}\approx 775$ MeV, and
\begin{equation}\label{hyb_hhg}
	(E_0-2m)^2=4 (2a) (2n+\ell+3/2).
\end{equation}
By comparing this formula with (\ref{masshl}), we see that this situation corresponds to a gluelump: A hybrid meson seen as a bound state of a gluon and a pointlike heavy meson. The string tension is $2a$ because it is the superposition of two fundamental strings. It can be computed that 
\begin{equation}
\left<r^2\right>=\frac{(2n+\ell+3/2)}{\beta}.	
\end{equation}
Consequently, thanks to definition (\ref{betadef3}), we have 
\begin{equation}\label{r2def}
	\left<r^2\right>=\frac{(2n+\ell+3/2)}{2a}\sqrt{\frac{\nu}{\mu}},
\end{equation}
and formula (\ref{hyb_hhg}) is thus equivalent to  
\begin{equation}
	E_0-2m=2(2a)\sqrt{\left<r^2\right>}.
\end{equation}
Half of the energy is given by the confinement and the other half by the kinetic energy of the gluon in agreement with the virial theorem. It also very interesting to compare formulas (\ref{regge_hyb}) and (\ref{hyb_hhg}). They both predict Regge trajectories depending on the color field excitation, but they differ in the interpretation they give to it. In the two-body case, this excitation is characterised by the quantum number $N$, defining the state of the string, while in the constituent gluon model, the excitation is represented by the gluon itself, and thus the Regge trajectory depends on its quantum numbers.   

In the general case, an analytical solution of Eq.~(\ref{condi}) can always be found, as shown in formula (\ref{musolu}) of Appendix \ref{eq3da} .
 It is worth noting that our procedure is not the same as the one developed in Ref.~\cite{Kalashnikova:2002tg}, where the AF were eliminated before computing the mass spectrum, and the resulting Hamiltonian was solved variationally with a numerical method.
\par The confinement interaction gives the correct behavior a large $R$. But, in order to be consistent in the region $R\ll\sqrt{a}$ we must add a OGE interaction between each pair of particles, i. e. 
\begin{equation}
	\Delta H = 2\frac{\kappa\, \alpha_S}{\sqrt{r^2+R^2/4}} + \frac{\kappa'\, \alpha_S}{R},
\end{equation}
with $\kappa = -3/2$ the color factor of (anti)quark-gluon pair, and $\kappa' = 1/6$ the color factor for the quark-antiquark pair~\cite{constg}. We used in $\Delta H$ the approximation (\ref{approx}), as for the unperturbed Hamiltonian \eqref{mainH}. The angular momentum is here a good quantum number even if we add the short-range interaction. This was not the case in \cite{Kalashnikova:2002tg}. As $\Delta H$ is computed perturbatively, the total energy reads, thanks to an usual approximation,
\begin{equation}\label{full_en}
	E -2m\approx E_0(\mu_0,\nu_0)-2m-\frac{3\alpha_S}{\sqrt{\left<r^2\right>+R^2/4}}+\frac{\alpha_S}{6R}.
\end{equation}
$\left\langle r^2\right\rangle$ is given by Eq.~(\ref{r2def}).
Let us notice that the Coulomb interaction for the $q\bar q$ system in a octet is repulsive. 

\section{The effective quark-antiquark potential}\label{eff_pot}

One of the observables in lattice QCD is the potential energy between the static quark-antiquark pair. It appears that there are several levels of potential energy, corresponding to different states of the gluon field \cite{Juge}. These excited states of the gluonic field are labeled by three quantum numbers. The first one is the projection $\Lambda$ of the total angular momentum $\vec J_g = \vec L_g + \vec S_g$ of the gluon on the $q\bar q$ axis. The capital greek letters $\Sigma,\Pi,\Delta,\ldots$ are used to indicate the states of $\Lambda = 0,1,2,\ldots$ respectively. The combined operations of the charge conjugation and the spatial inversion of the quark and of the antiquark is also a symmetry. Its eigenvalue is denoted by $\eta_{CP}$. States with $\eta_{CP}=1 (-1)$ are denoted by the subscripts $g$ ($u$). There is a additional label for the $\Sigma$ states: $\Sigma$ states which are even (odd) under a reflexion in a plane containing the $q\bar q$ axis are denoted by a superscript $+$ ($-$). All these different states have been observed in Ref.~\cite{Juge2}.

\begin{figure}
\resizebox{0.5\textwidth}{!}{%
  \includegraphics*{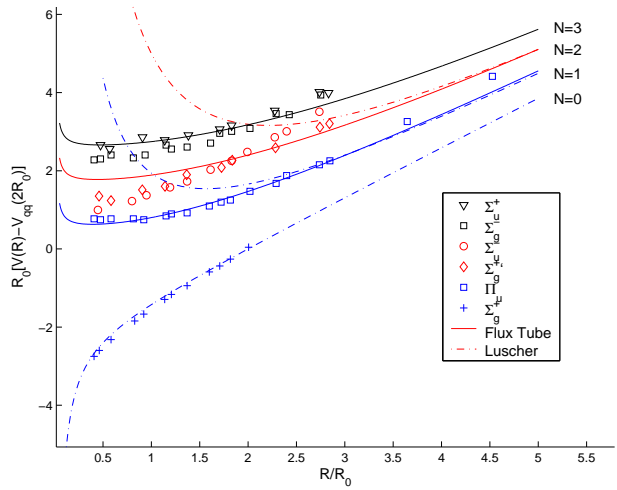}
}
\caption{Comparison between lattice QCD calculations (symbols) from Ref.~\cite{Juge}, and two-body models like the excited flux tube (solid lines) from Eq.~\eqref{pot_ols} and the L\"uscher term (dot-dashed lines) from \eqref{pot_lu}. All the potentials are plotted in terms of the lattice scale $R_0 = 2.5$ GeV$^{-1}$ and are shifted by an overall amount $V_{q\bar q}(2R_0)$. The parameters $a=0.2$ GeV$^{2}$ and $\alpha=0.3$ are fitted  on the lattice ground state $\Sigma_g^+$.}
\label{fig:nambu}
\end{figure}

In the excited flux tube picture, the glue states with $N=0$ and $N=1$ are uniquely the ground state $\Sigma_g^+$ and $\Pi_u$ respectively. For $N>1$, the flux tube can be excited in $\Lambda = 0,1,\ldots,N$ states \cite{Allen:1998wp}. The CP value is given by $\eta_{CP}=(-1)^N$. Potentials (\ref{pot_lu}) and (\ref{pot_ols}) are compared to the lattice data in Fig.~\ref{fig:nambu} for the lowest states. As remarked in Ref.~\cite{Allen:1998wp}, the particular string potential (\ref{pot_ols}) fit the excited levels with a good accuracy, while the L\"uscher potential \eqref{pot_lu} reproduces very well the ground state but diverges too fast at small $R$ for $N>0$.
\begin{figure}
\resizebox{0.5\textwidth}{!}{%
 \includegraphics{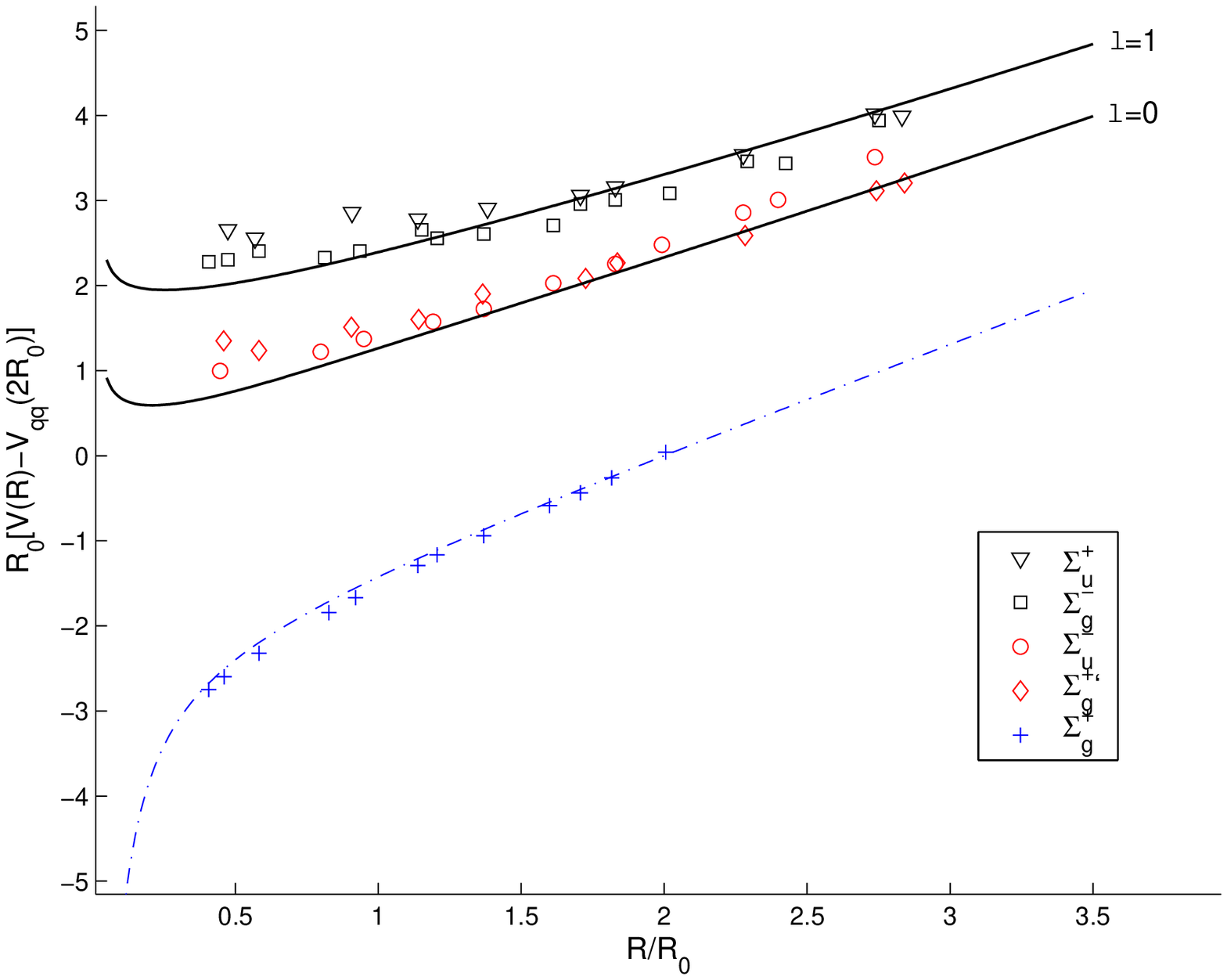}
  }
\caption{Same as Fig.~\ref{fig:nambu}, but the solid lines now represent our $q\bar qg$ mass formula (\ref{full_en}). The upper curve is the P-wave ($\ell=1$), and the lower curve is the S-wave ($\ell=0$). The ground state is shown only for comparison, and is a funnel potential with the parameters of Fig.~\ref{fig:nambu}  }
\label{fig:sigma_nous}
\end{figure}

\begin{figure}[t]
\resizebox{0.5\textwidth}{!}{%
  \includegraphics*{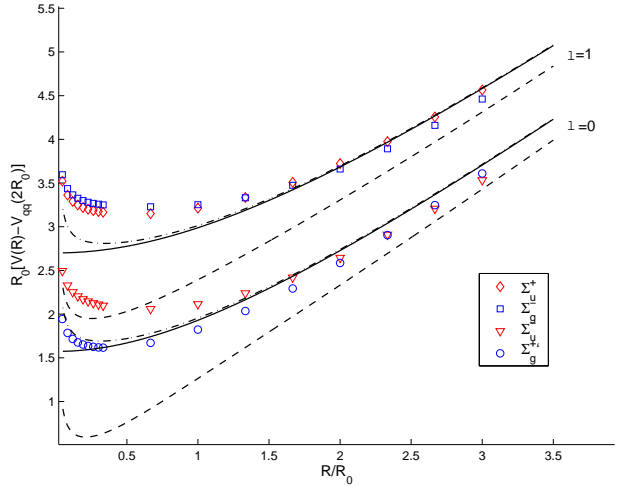}
  }
\caption{Comparison between the constituent gluon model of Ref.~\cite{Szczepaniak:2006nx} (symbols), and our $q\bar q g$ model without short range interaction (solid lines), with $q\bar q$ OGE term (dot-dashed lines) and with the total OGE term (dashed lines). Parameters are those of Fig.~\ref{fig:nambu}.}
\label{fig:sigma_1}
\end{figure}

In our constituent gluon model, defined by the mass formula (\ref{full_en}), we will assume for the gluon that $n=0$. $\ell$ is then the only relevant quantum number. Since the $q\bar q$ system and the gluon have both a negative intrinsic parity, the parity of the states is the space parity $(-1)^\ell$. The charge conjugation on the gluon give a $-1$ factor, and a $(-1)^{S_{q\bar q}}$ factor for the quark-antiquark pair. The value of $\eta_{CP}$ is thus $(-1)^{\ell+S_{q\bar q}+1}$ and can give either $g$ or $u$ states. But, in our case, we made a strong symmetry assumption, and considered that the gluon was always located in the symmetry plane of the $q\bar q$ pair. Thus, the only value we can reproduce is $\Lambda=0$ with a positive subscript, since it is the only value in agreement with the symmetry of our wave function. With these considerations, we can conclude that the states described by our model are $\Sigma^+$ states. The spin effect being weaker for maximum spin states \cite{Mathieu:2005wc}, we should better reproduce the states with $S_{q\bar q}=1$. More precisely, we get the $\Sigma^+_g$ for the S-wave ($\ell=0$), and the $\Sigma^+_u$ for the P-wave ($\ell=1$). 
The comparison between the lattice $\Sigma$ states and the model for $\ell=0,1$ with all short-range interactions is shown in Fig. \ref{fig:sigma_nous}. Although our formula is simple and approximated, one can see that it fits with a good agreement the lattice data. Let us note that, following our discussion on the quantum numbers $C$ and $P$, the P-wave corresponds to a $1^{-+}$ hybrid meson.
	\begin{figure}[t]\label{diagrams}
	\resizebox{0.45\textwidth}{!}{%
	  \includegraphics*{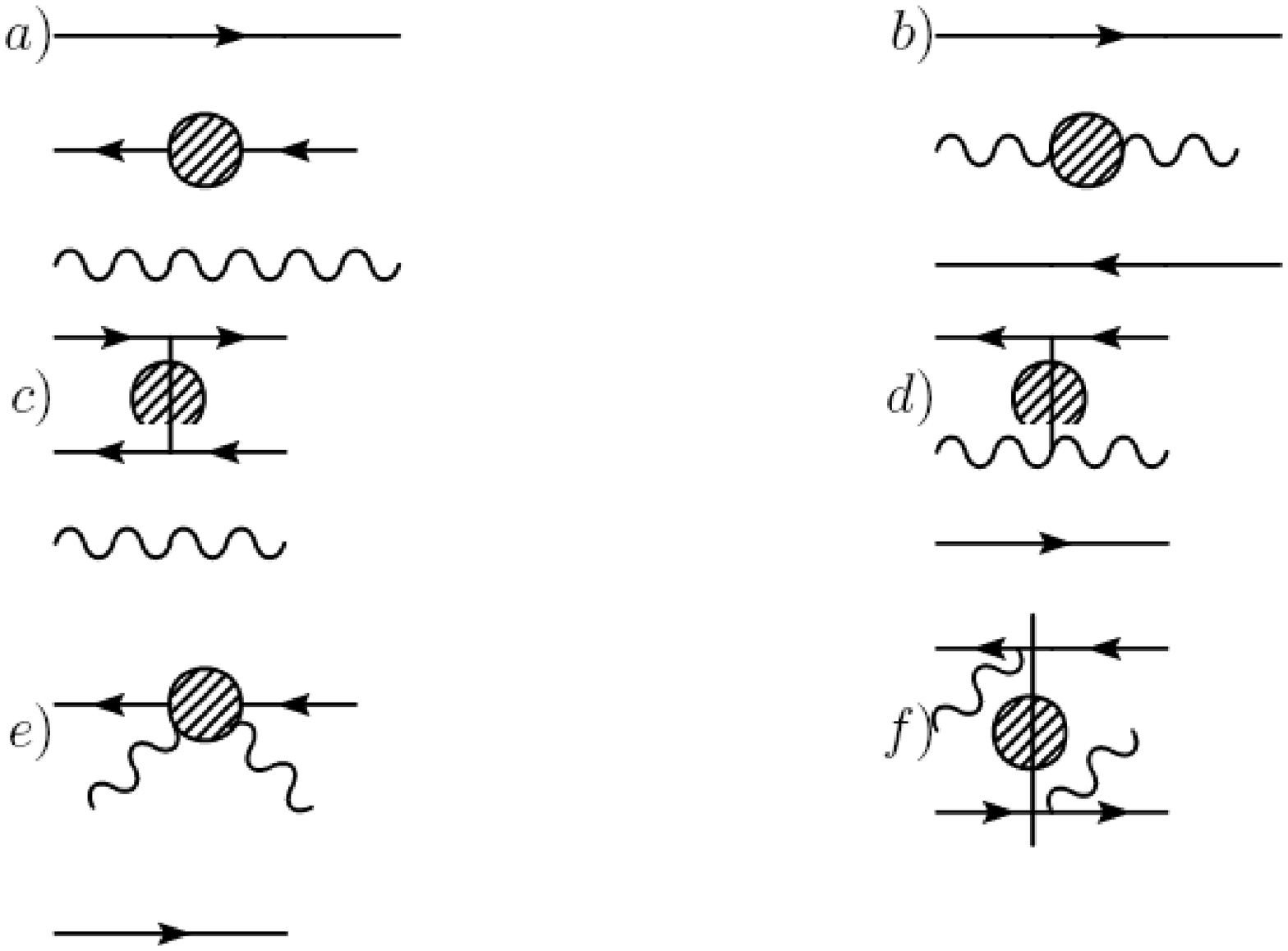}
	  }
	
	\end{figure}

Other recent studies of hybrid mesons with a constituent gluon can also be mentioned \cite{Szczepaniak:2005xi,Szczepaniak:2006nx,Swanson:1998kx}. In all these works, based on QCD in the Coulomb gauge and the quasi-particle representation, the Hamiltonian of the system was solved variationally by a numerical method. In a first work \cite{Swanson:1998kx}, the level ordering of the states at short distance disagreed with lattice results. For example, the $\Pi_g$ state was found below the $\Pi_u$ at short distance. Recently, the authors added new interactions, listed in the diagrams below \cite{Szczepaniak:2006nx}. $a$ and $b$ are the self-energies of the quarks and the gluon respectively, and $b$ and $c$ are the quark-antiquark, and (anti)quark-gluon interactions, which are responsible for both the confinement and the short-range term. These four diagrams were the only ones considered in Ref.~\cite{Swanson:1998kx}. The new diagrams are interpreted as particular three-body interactions, and are represented by $e$ and $f$ \cite{Szczepaniak:2006nx}. We present in Fig.~\ref{fig:sigma_1} the comparison between our analytic model and the numerical data of Ref.~\cite{Szczepaniak:2006nx}. The solid lines represent the energy \eqref{energ_h1} of the Hamiltonian with only the confinement for $\ell=0,1$. The dashed-dotted and dashed lines are the energies with the $q\bar q$ Coulomb repulsive interaction and with all the short-range interactions respectively. Our results are somewhat different at small $R$, but agree in the confining part. It can also be checked that the full energy (dashed lines) better fits the plot of the first study \cite{Swanson:1998kx}. This can be understood by noting that our model only includes the confinement and the coulomb term coming from OGE, thus the diagrams $b$ and $c$. We neglected the self-energy $a$ and $b$, but it should not be a dominant contribution in potential models. Indeed, in the case of heavy quarks, it is very small \cite{qse}. Moreover it is even argued in some approaches that the gluonic self-energy is vanishing \cite{qse2,qse3}. However, it seems that the three-body interactions $e$ and $f$, that we did not took into account, bring a relevant contribution to the total energy. This could be clarified in future works.

\section{Hybrid spectrum}\label{spectrum}

Masses of heavy hybrid mesons can be derived from Eq.~(\ref{full_en}) by taking the miminum of the energy respect to $R$. We find a that $E-2m=1.700$ GeV is the minimum value for the P-wave. As we argued in the previous section, this P-wave represents a $1^{-+}$ hybrid. Many lattice studies have been devoted to such exotic mesons, and we can compare our predictions with these results. This is done in Tab.~\ref{tab:spec}. For the minimum of the S-wave, we find the value $E-2m=1.16$ GeV. This gives a $c\bar cg$ mass around $3.66$~GeV for the $1^{++}$ state in disagreement with the lattice result $7.315$ GeV of ref. \cite{Luo:2005zg}. Nevertheless, our value is close to the experimal state X(3872) \cite{Choi:2003ue} which could be interpreted as a hybrid meson but also as a meson-antimeson molecule \cite{Petrov:2006zq}.

\begin{table}[ht]
	\centering
		\begin{tabular}{l|c|c|c}
	$q$	& $m_q$  & $M_{q\bar q g}$  & $M_{\rm{lat}}$ \\
		\hline
$n$ &	$0.005\pm0.003$ & $1.710\pm0.006$ & $1.740\pm0.240$ \cite{Hed05} \\
$s$	& $0.105\pm0.025$ & $1.910\pm0.050$ &  $2.100\pm0.120$ \cite{Bern03} \\
$c$ & $1.250\pm0.100$ & $4.200\pm0.200$ &  $4.405\pm0.038$ \cite{Liu06}\\
$b$ & $4.500\pm0.400$ & $10.700\pm0.800$ &	$10.977\pm0.123$ \cite{Mank01}\\
		\end{tabular}
	\caption{$1^{-+}$ hybrids masses in our constituent gluon model, $M_{q\bar q g}$, compared to lattice QCD computations, $M_{\rm{lat}}$. Masses are in GeV, and the quark masses $m_q$ are taken from the Particle Data Group \cite{pdg}. The errors on our results are simply computed from the Particle Data Group errors on the quark masses. }
	\label{tab:spec}
\end{table}

Obviously, our mass formula is only valid for the heavy quarks, namely $c$ and $b$. But we see in Tab.~\ref{tab:spec} that these masses are even in good agreement with recent lattice results concerning light hybrids $n\bar ng$ and $s\bar sg$. This could be an indication that the dominant degree of freedom in a hybrid meson is the one coming from the constituent gluon, the quarks mainly bringing a rest mass term to the total mass. 
However, if we compare our results to a previous work based on QCD in Coulomb gauge \cite{llanes}, we find that the agreement is good for the charmed hybrid, where the dynamical effects are small, but rather bad in the light sector. This shows that the dynamical effects of the quarks, following the way they are taken into account, considerably affect the hybrid spectrum. Such effects should thus be studied with more details in future works. Let us finally remark that, in our model, we found $E-2m=1.700$ GeV to be a constant, which gives the energy coming from the constituent gluon in the P-wave. This is close to the exotic meson $\pi_1(1600)$ \cite{Ada98}. In the excited flux tube model, the equivalent energy of the string fluctuation is $\sqrt{2\pi a N}$, with $N=3$, that is $1.942$ GeV, a higher value than with our constituent gluon approach. In the Coulomb gauge model, $E-2m=1.825$ GeV in the charmed sector, in agreement with our result. 

\section{Summary of the results}\label{conclu}
In this work, we studied the hybrid mesons in the framework of potential models. In particular, we applied the auxiliary fields technique to obtain analytical mass formula and wave functions of the spinless Salpeter Hamiltonian with linear confinement. Firstly, we showed that in the well-known case of mesons, the auxiliary fields allow to get easily mass formula whose features are qualitatively in agreement with experiment: The Regge trajectories for mesons formed of two light quarks or one heavy and one light quark are correctly predicted. 

The simplest way to study hybrid mesons is to work in the excited flux tube framework. It is based on the idea that the flux tube (a Nambu-Goto string) linking the quark and the antiquark is not in its ground state, but in an excited one. Since the work of Isgur and Paton \cite{Isgur:1984bm}, the relativistic vibrating string models were widely discussed in the litterature. In particular, it was shown in Ref.~\cite{luscher} that the first order correction of the excited flux tube was a universal term given by $\pi(N-1/12)/r$. In Ref.~\cite{Allen:1998wp}, it was suggested from string theory that the confinement potential should be modified to $\sqrt{a^2r^2+2\pi a N}$, this formula giving good results when compared to lattice calculations \cite{Juge}. As in this approach we always deal with a two-body problem, the mass spectra can again be easily computed. In particular, we showed that in a hybrid meson formed of two heavy quarks of mass $m$, one should have $(M_{hh}-2m)^2\propto N$. 

We also considered an other picture, which assimilates the hybrid meson to a three-body quark-antiquark-constituent gluon bound state \cite{constg}. In the case of two static quarks of the same mass, we computed the mass spectrum of the corresponding hybrid meson. As a result, we have been able to find an analytic expression for the interquark potential in terms of the quantum numbers of the gluon. Although our model was very simple, it correctly leads to the gluelump spectrum if the quark-antiquark separation is zero, and it reproduces rather well the lattice data which can be described within our assumptions. We was actually only able to reproduce the $\Sigma^+$ curves, but the other states can also be described following the quantum numbers of the gluon \cite{Kalashnikova:2002tg}. In the gluelump case moreover, one can observe that $(M_{hh}-2m)\propto\ell$. This illustrates the similarity between the excited flux tube and the constituent gluon formalisms, but the degrees of freedom are different: The excitation number $N$ is replaced by the gluon orbital momentum $\ell$. Other works studied the picture of a constituent gluon with two static quarks \cite{Szczepaniak:2006nx,Swanson:1998kx}. An interesting point is that our curves correctly fit previous works \cite{Swanson:1998kx}, but slightly disagree with more recent references, where three-body interactions, which we neglected, are taken into account \cite{Szczepaniak:2006nx}. The relevance of such interactions in potential models could be studied in the future. We have also shown that our model could well reproduce lattice QCD data, and in particular the P-wave, corresponding to the $1^{-+}$ hybrids. The hybrid masses we obtained are roughly in agreement with lattice, and also compatible with the $\pi_1(1600)$. However, the mass we get for the $n\bar n$ hybrid disagrees with a previous study based on QCD in the Coulomb gauge, because of the dynamical effects of the quarks, which are rather strong in this last approach.  

Our study only aims to understand in an intuitive way qualitative features of hybrid mesons. In future works, the main challenge will be to compute precisely the various interactions coming from the dynamics and spin of the particles, which we neglected here. To do this, we think that a model with constituent gluons is more interesting, because a constituent gluon has its well defined color, spin and interactions with quarks. On the contrary, the only clear characteristic of an excited flux tube is the quantum number $N$. We leave such detailed studies of the spin effects in hybrids for future work. 

The authors would like to thank the FNRS Belgium and IISN for financial support. We are grateful to Dr. Fabian Brau and Dr. Claude Semay for advices and useful discussions, and to Dr. Adam. Szczepaniak for providing us the data of Refs.~\cite{Szczepaniak:2006nx,Juge}. 

\begin{appendix}
\section{Influence of the auxiliary fields on the mass spectrum}\label{AFM}

\par Let us suppose that the Hamiltonian of our problem is of the form
\begin{equation}\label{hamA1}
	H_0=\sum^N_{i=1}A_i,
\end{equation}
 where $A_i$ are some operators. The eigenvalues $E_0$ and eigenstates $\left|\psi_0\right\rangle$ are assumed to be known. We can introduce $k$ AF, denoted as $\phi_i$, to obtain a new Hamiltonian $H_k$,
\begin{equation}\label{hamA2}
	H_k=\sum^k_{i=1}\left(\frac{A^2_i}{2\phi_i}+\frac{\phi_i}{2}\right)+\sum^N_{j=k+1}A_j.
\end{equation}
As an illustration, one can compare Hamiltonians (\ref{hamA1}) and (\ref{hamA2}) to the SSH (\ref{ham1}) and (\ref{ham2}).
$H_k$ is equivalent to $H_0$ if the AF are directly eliminated as operators. However, our method is to consider that the $\phi_i$ are real numbers, in order to simplify the calculations. Let us suppose that we know $E_k$ and $\left|\psi_k\right\rangle$ the solutions of the eigenequation defined by $H_k$. Then, clearly,
\begin{eqnarray}
	\left\langle \psi_k\right|(A_i&-&\phi_i)^2\left|\psi_k\right\rangle\geq0\nonumber\\
	 &&\Rightarrow\ \left\langle \psi_k\right|\frac{A^2_i}{2\phi_i}+\frac{\phi_i}{2}\left|\psi_k\right\rangle\geq\left\langle \psi_k\right|A_i\left|\psi_k\right\rangle .
\end{eqnarray}
This implies that 
\begin{equation}
	E_k=\left\langle \psi_k\right|H_k\left|\psi_k\right\rangle \geq \left\langle \psi_k\right|H_0\left|\psi_k\right\rangle \geq E_0.
\end{equation}
Moreover, the same argument immediately allows to show that 
\begin{eqnarray}
	E_{k+1}&\geq& E_{k}-\frac{1}{2\phi_{k+1}}\left\langle \psi_{k+1}\right|(A_{k+1}-\phi_{k+1})^2\left|\psi_{k+1}\right\rangle\nonumber\\
	&\geq& E_k,
\end{eqnarray}
the only condition being that the $\phi_j$ are positive. This is always the case in the situations we treated here, since the AF are interpreted as effective quark mass and string energy. Thus, we have proved that the masses obtained with the AF technique are an upper bound of the true masses, as already shown in Ref.~\cite{Fab3}, but also that this upper bound is less and less strong when the number of AF increases. As an example, we can note that the Regge slope coming from the full two-body SSH without AF is around $2\pi a$ \cite{Ols94}. With only one AF, it has been shown in Ref.~\cite{qse} that the slope was roughly equal to $7a$. Here, with two AF, we observed a slope given by $8a$. So, the slope increases with the number of AF. Instead of the mass formula (\ref{mass2_4}), we should actually write the following inequality:
\begin{equation}
	M^2_{ll}\leq8a(2n+\ell+3/2).
\end{equation}

\section{Cardan method for third degree equations}\label{eq3da}
In this section, we give the explicit solution of Eq.~\eqref{condi}. With $\mu^2_0=X$, this condition can be rewritten as
\begin{equation}\label{condi2}
	X^3 + \frac{k^2}{a^2R^2}X^2-\frac{k^4}{a^2R^2}=0.
\end{equation}
Since $\mu_0$ is interpreted as an effective quark mass, $X$ has to be a positive real number. The corresponding solution of Eq.~(\ref{condi2}) is
\begin{equation}
	X_0= \sqrt[3]{V + \sqrt{Q^3+V^2}} +\sqrt[3]{V - \sqrt{Q^3+V^2}},
\end{equation}
with 
\begin{eqnarray}
	Q &=& -\frac{k^2}{9a^2R^2}\\
	V &=& \frac{k^4}{2a^2R^2} - \frac{1}{27}\left(\frac{k^2}{a^2R^2}\right)^3.
\end{eqnarray}
Then, we have simply
\begin{equation}\label{musolu}
	\mu_0=+\sqrt{X_0}.
\end{equation}

\end{appendix}

\end{document}